\documentclass[aps,prb,floatfix,preprint]{revtex4}
\usepackage{amssymb}
\usepackage{amsmath}
\usepackage{epsfig}
\usepackage{graphicx}
\usepackage{graphics}
\usepackage[english]{babel}

\begin{document}

\title{Hydrodynamic model for relaxation of optically injected currents in
quantum wells
\footnote{A slightly shorter version published in Appl. Phys. Lett. \textbf{91}, 232113 (2007) }}
\author{R.M. Abrarov, E.~Ya.~Sherman, and J.~E.~Sipe}

\begin{abstract}
We use a hydrodynamic model to describe the relaxation of optically injected
currents in quantum wells on a picosecond time scale, numerically solving
the continuity and velocity evolution equations with the Hermite-Gaussian
functions employed as a basis. The interplay of the long-range Coulomb
forces and nonlinearity in the equations of motion leads to rather complex
patterns of the calculated charge and current densities. We find that the time
dependence of even  the first moment of the electron density is sensitive to this
complex evolution. 
\end{abstract}

\affiliation{Department of Physics and Institute for Optical Sciences, University of
Toronto, 60 St. George Street, Toronto, M5S 1A7, Canada}
\maketitle

The dynamics of hot electron currents, which determines the distance that
injected carriers can propagate and the charge density patterns that can
form, is important for the design of various semiconductor devices,
including transistors, charge-coupled light detectors, cascade lasers and
light emitting diodes \cite{Chicago}. The full dynamics of
systems strongly out of equilibrium requires a complicated analysis based
either on extended numerical modeling employing Monte Carlo simulations 
\cite{MonteCarlo}, or on quantum versions of the kinetic equations\cite{Wu06,Duc06,Rumyantsev04,Kuhn04}.
For two-dimensional (2D), systems quantum kinetic approaches
are numerically demanding and often unfeasible.

The direct injection of strong electrical and spin currents 
\cite{Bhat00} is possible using the interference of one-photon 
(frequency $2\omega $) and two-photon (frequency $\omega$) absorption. The high speed
of the injected electrons, $v\approx 1000$ km/s, is
determined by the excess photon energy $2\hbar\omega-E_{g},$ where $E_{g}$
is the band gap for the quantum well (QW). In an experiment injecting
electrical current, coherent
control is achieved by adjusting the relative phase parameter $\Delta\phi =2\phi
_{\omega }-\phi _{2\omega}$ of the two beams, where $\phi _{\omega }$ $%
\left( \phi _{2\omega }\right) $ is the phase of the beam at $\omega $ $%
\left( 2\omega \right).$ The resulting current density is approximately given by $Nv\sin(\Delta\phi),$ 
where $N$ is the concentration of optically injected carriers.

In experiments done on multiple quantum well (MQW) samples, with the beams
incident along the growth direction, these injected lateral currents 
(see Fig.1) have been detected
using various techniques \cite{Stevens02,Hubner03}.    The
QWs are well-separated and unbiased, so electrons do not have
sufficient energy either to tunnel through the barriers or overcome them by
thermal activation. For this reason no dynamics along the "vertical" growth direction
is possible, and in these systems we do not explore
the rich nonlinear behavior seen in vertical transport in
superlattices.\cite{Bonilla05}. 
Instead, our subject is only the lateral dynamics of the carriers
subsequent to their injection.

Nonetheless, the analysis of this lateral dynamics is more complicated than
in many transport problems because of the largely inhomogeneous long-range
Coulomb field, leading to complicated space-charge effects. As the electrons
and holes move away from each other, the characteristic range of
inhomogeneity is on the order of the size of the electron and hole puddles
themselves. A qualitative approach to this problem is to adopt a
"rigid spot" approximation, which reduces the problem to
the motion of two coupled and damped linear oscillators, representing the
centers of the electron and hole puddles, each moving with 
a uniform velocity and thus exactly preserving its shape.\cite{Sherman06} Here the
entire dynamics is described by four parameters: displacements of electron
and hole spots and their velocities.

An important question is to what extent this "rigid spot" model can
describe the dynamics, as Coulomb and other
concentration-dependent forces come into play when the puddles
separate. Typically, the change in carrier densities at any point is
small, since the relaxation and Coulomb interaction prevent any large separation of the
centers of charge. However, can the spatial dependence of this small
change depart significantly from what is predicted in the "rigid spot" model,
leading to a picture involving more generally  "distorted puddles" as more appropriate ?

Answering this question is important for the ultimate design of
numerical schemes to efficiently explore the 
system dynamics. 
What is needed initially is a practical model and calculation
procedure that captures the essential physics, is transparent enough to
allow extraction of the important features of the dynamics, and
reasonable enough to allow comparison with experimental results, at
least at a semiquantitative level. Here we propose a phenomenological
hydrodynamic model for the dynamics of optically injected charge currents
subject to space-charge effects. The advantage of the hydrodynamic approach
is that it is insensitive to the details of the carriers' distribution
functions, with the dynamics described by a coupled system of partial
differential equations for the concentrations and velocities. 
The collisions between carriers, and interaction between the
carriers and the environment, are modeled by a set of characteristic times
determined by the experimental conditions. 

As an example calculation we consider a MQW
structure (Fig. 1), assuming the excitation and subsequent dynamics in each
quantum well is the same; within our model we perform a full 2D calculation
of the carriers' motion. The $\omega $- and $2\omega $-beams produce initial
distribution of electron and hole densities and velocities which then evolve
in time and space.  The initial distribution of carriers in
each single QW is 
$N_{e,h}^{(s)}\left(\mathbf{r},t=0\right)\equiv N_{s}\exp\left(-r^{2}/2\Lambda^{2}\right)$, 
where $\Lambda$ is governed by the beam sizes (see caption of Fig.1). 
The total concentration 
$N_{e,h}\left( \mathbf{r},t\right)=qN_{e,h}^{(s)}\left(\mathbf{r},t\right)$, 
where $\mathbf{r}=(x,y)$ is the 2D coordinate, and  $q$ is 
the total number of single QWs, neglecting the distance
from the first to the last single QW in the structure compared to $\Lambda$.
The spin indices are omitted since here the electrons are
assumed to be unpolarized. The $\left(\mathbf{r},t\right)$ 
arguments will be omitted below for brevity.

For simplicity, in this letter we present results for the dynamics of
electrons assuming that the holes are infinitely heavy;  we
will include hole dynamics in a later publication. The equations describing the motion of
carriers consist of the continuity, and momentum and energy evolution sets.
In the effective mass approximation, the analysis of the dynamics shows that
a first description can be provided even without taking the energy
relaxation into account. For this reason we restrict ourselves to the first
two sets of variables. The continuity and the evolution of the 
local mean velocity $\mathbf{u}\left(\mathbf{r},t\right)$ of electrons, are
described by:
\begin{eqnarray}
&&\frac{\partial N_{e}}{\partial t}+\nabla \left( N_{e}\mathbf{u}\right) =0, \\
&&\frac{\partial \mathbf{u}}{\partial t}+\left( \mathbf{u}\mathbf{\cdot \nabla 
}\right) \mathbf{u}=-\frac{e\mathbf{E}}{m_{e}}-\frac{\mathbf{u}}{\tau _{eh}}%
\frac{N_{h}}{N_{0}}
-\frac{\mathbf{u}}{\tau _{e}},
\end{eqnarray}%
where $\mathbf{E}=\mathbf{E}^{e}+\mathbf{E}^{h}$ is the macroscopic Coulomb
field produced by electrons and holes, and $e$ is the elementary charge.
Here $m_{e}$ is the electron effective mass, $\tau _{eh}$ describes
momentum-conserving drag due to the Coulomb forces at electron-hole
collisions \cite{Hopfel} and can be weakly concentration-dependent itself \cite%
{Zhao07}, $\tau _{e}$ is the relaxation time for electrons
due to external factors, such as phonons and impurities leading to the
relaxation of the total momentum, and  $N_{0}=qN_{s}$.  
In addition to the explicit time scales $\tau _{e}$ and $\tau _{eh}$
a third important time scale in the problem is $\Omega_{\rm pl}^{-1}$, 
where 
$\Omega_{\rm pl}=$ $\left(\pi /2\right)^{3/4}\sqrt{N_{0}e^{2}/\varepsilon m_{e}\sqrt{2}\Lambda}$ 
is the characteristic
two-dimensional plasma frequency\cite{Sherman06} ($\varepsilon$ is the dielectric
constant); it characterizes the
strength of the Coulomb interaction. In our sample calculations
we take $\tau_{e}=80$ fs and $\tau_{eh}=150$ fs;  a
typical value $N_{s}=10^{11}$ cm$^{-2}$ with $q=1$ and our choice of 
$\Lambda=1$ $\mu$m leads to $\Omega_{\rm pl}^{-1}\approx 1.4$ ps.
These times agree with the set of parameters of Duc {\it et  al.} [\onlinecite{Duc06}] and
are shorter than those for vertical transport \cite{Weber06} due to the higher 
dimensionality of carrier motion here. 

\begin{figure}[t]
\includegraphics[width=0.4\columnwidth]{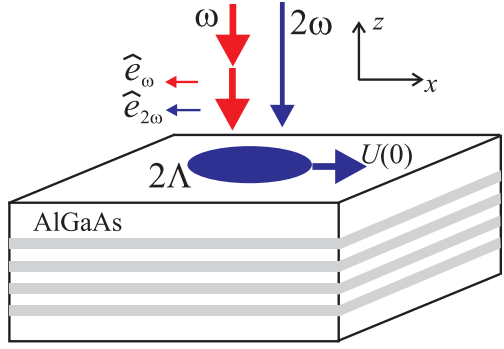}
\caption{(Color online) A sketch of the optical current injection scenario.
The MQW structure with $q=4$ (gray bands for GaAs) is shown.
Vectors $\widehat{e}_{\protect\omega (2\protect\omega )}$ correspond to the
polarization of the photons. The injected charge density follows the
profiles of the $2\omega$-beam and the $square$ of the  $\omega$-beam 
intensity, which are the same due to the $2\omega$ generation procedure
(frequency doubling in nonlinear crystal). $U(0)$ is the initial macroscopic speed of the
electron puddle.}
\end{figure}

To reduce the system of partial differential equations, we introduce a full
basis set in the form of the harmonic oscillator wave functions 
\begin{equation}
\Psi _{n_{1},n_{2}}\left(\mathbf{r}\right) =\psi _{n_{1}}(x)\psi
_{n_{2}}(y),\qquad \psi _{n}(x)=\frac{H_{n}\left( x/\Lambda \right) }{\pi
^{1/4}\sqrt{n!2^{n}}}e^{-x^{2}/2\Lambda ^{2}},
\end{equation}%
where $H_{n}\left(x/\Lambda\right)$ is the Hermite polynomial of the $n-$%
th order. The key point of our approach is the expansion of the possible
solutions in a finite basis in the form:
\begin{equation}
N_{e} =\sum_{n_{1},n_{2}}^{n_{\max }}N_{n_{1},n_{2}}^{e}(t)\Psi
_{n_{1},n_{2}}\left(\mathbf{r}\right);  \qquad
u_{i} =\sum_{n_{1},n_{2}}^{n_{\max }}u_{n_{1},n_{2}}^{(i)}(t)\Psi
_{n_{1},n_{2}}\left(\mathbf{r}\right) +U_{i}\left( t\right) .
\end{equation}
Here $i=x,y$ is the Cartesian index. To improve the convergence, we include
a known function of time $U_{i}\left( t\right) $ in the right-hand side of
Eq.(4). This function is obtained by
solving the linear equations of motion in the rigid spot approximation \cite%
{Sherman06}. In the geometry considered here, $U_{y}\left( t\right) =0$,
and, therefore, we drop the Cartesian index of $U_{x}\left( t\right) $. The
electric fields at the point $\mathbf{r}$ are 
\begin{equation}
\mathbf{E}^{e,h}\left(\mathbf{r},t\right) =\mp \frac{e}{\varepsilon }%
\sum_{n_{1},n_{2}}^{n_{\mathrm{max}}}N_{n_{1},n_{2}}^{e,h}(t)
\int 
\Psi_{n_{1},n_{2}}\left(\widetilde{\mathbf{r}}\right)\frac{\mathbf{r}-%
\widetilde{\mathbf{r}}}{\left\vert \mathbf{r}-\widetilde{\mathbf{r}}%
\right\vert ^{3}}d^{2}\widetilde{r},
\end{equation}%
where the upper (lower) sign
corresponds to electrons (holes). If a QW is located close to the
semiconductor-air interface, the role of the image charges which can be
taken into account with the Green function technique \cite{Sipe81}, increases
the field by a factor of two. Since
$\Lambda $ is typically much larger than the distance
between the QWs, within our assumption that the excitation of all
the wells is the same, the electric field at a point $\mathbf{r}$ in each
well and the subsequent dynamics is the same. 

The state and the motion of the electron spot is then fully described by a $%
3\times \left( n_{\max }+1\right) ^{2}-$component vector $S_{\alpha },$
where $\alpha $ is a two-component index corresponding to indices $n_{1}$
and $n_{2}.$ By projecting the equations (1),(2) and (5) on the set of $\Psi
_{n_{1},n_{2}}\left( \mathbf{r}\right) $, we obtain the system of ordinary
nonlinear differential equations in the form 
\begin{equation}
\frac{dS_{\alpha }}{dt}=\sum_{\eta,\mu}\mathcal{C}\left( \alpha ;\eta ,\mu
\right) S_{\eta }S_{\mu }-\sum_{\zeta}{\Gamma }\left( \alpha ;\zeta
\right) S_{\zeta }.
\end{equation}%
Here the matrix $\mathcal{C}\left( \alpha ;\eta ,\mu \right) $ is determined
by the spatial dependence of concentration and velocity, and the
electron-hole drag, while the matrix ${\Gamma }\left( \alpha ;\zeta \right) $
depends on the Coulomb forces and momentum relaxation times. The full
matrices will be presented elsewhere.

The initial conditions are given by: (i) $N_{0,0}^{e}(0)=\sqrt{\pi }N_{0},\;N_{n_{1},n_{2}}^{e}(0)=0$ (if $n_{1}>0$ or $n_{2}>0$), and (ii) $%
u_{n_{1},n_{2}}^{(i)e}(0)=0.$ Conditions (i) correspond to the initial
injection of density in the Gaussian mode $\Psi_{0,0}\left( \mathbf{r}\right) $
only. Due to the symmetry of the problem $N_{n_{1},2m+1}^{e}(t)$, $%
u_{n_{1},2m+1}^{(x)}(t)$ and $u_{n_{1},2m}^{(y)}(t)$ will remain zero. The
initial speed of the electron spot $U(0)$ is determined \cite{Bhat00} by the
photon excess energy and $\Delta\phi$. 

At very short times after the injection the carriers propagate
ballistically, and then the motion becomes diffusive and influenced by the
Coulomb forces. A simple characterization of the motion of the charge is
given by the first moment of the concentration,%
\begin{equation}
\left\langle x(t)\right\rangle =\frac{1}{N_{\rm t}}{\int xN_{e}\left( \mathbf{r},t\right)
d^{2}r},\quad 
N_{\mathrm{t}}=\pi N_{0}\Lambda ^{2},
\label{firstmoment}
\end{equation}%
where $N_{\mathrm{t}}$ is the total number of
injected electrons. The relaxation and Coulomb interaction guarantee that 
$\xi\equiv\left\langle x(t)\right\rangle/\Lambda $ is always small in our systems.
Were the rigid spot approximation valid, the amplitude of the nonzero higher
terms $N_{n,0}^{e}$ would drop off as $\xi^{n}$.  The main
qualitative result of our investigations is that \textit{this is often not
so} for reasonable excitation parameters.  Instead, the
concentration-dependent Coulomb forces and drag lead to a deformation of the
spot and, therefore, to an increase in the number of $\Psi _{n_{1},n_{2}}$ terms
with non-negligible amplitude in the sums, and to a complicated spot
structure.  In the course of time, a few of the 
$N_{n_{1},n_{2}}^{e}(t)$  components can become of the same order of magnitude, although
all small compared to $N_{0,0}(0)$ due to a relatively small spot
displacement $\xi\ll 1$. The number of
higher terms that are important depends mainly on the structure of the $%
\mathcal{C}\left( \alpha ;\eta ,\mu \right) $-matrix, with their relative
contribution dependent on the parameters $\tau _{e}\Omega _{\mathrm{pl}}$
and $\tau _{e}/\tau _{eh}$.  In contrast to the scattering by phonons and
disorder, which stabilizes the spot motion and does not lead to its
distortion, the role of electron-hole collisions is two-fold. On one
hand, it stops the motion of the spot as a whole, on the other hand, it
generates higher components in the velocity pattern due to nonuniform
velocity evolution. The nonuniform velocity distribution produced by the
Coulomb forces and the drag then leads to a nonuniform density distribution
due to continuity. So the joint effects of the Coulomb forces and the drag
lead to a "buckling" of the electron
spot along the direction of its velocity. The buckling
becomes more pronounced as the space-charge effects,
characterized by the parameter $\tau _{e}\Omega _{\mathrm{pl}}$, increase. 

\begin{figure}[t]
\includegraphics[width=0.55\columnwidth]{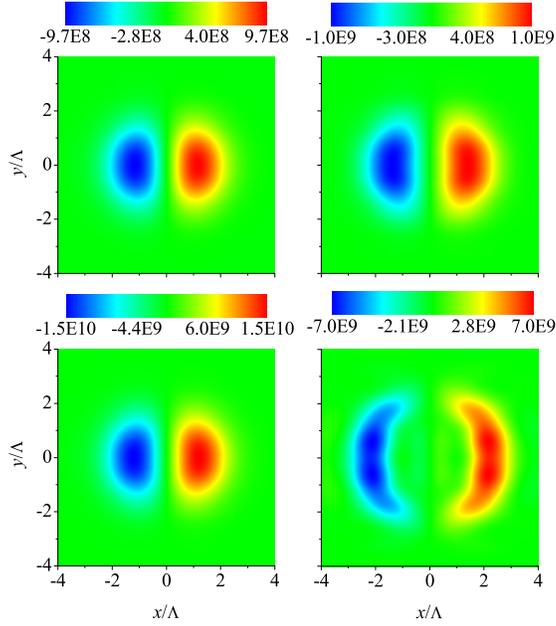}
\caption{(Color online) The change in the electron density distribution
compared to the injected density at $t=80$ fs (left column) and $t=1.5$ ps
(right column). The MQW parameters are $q=1$ (upper row) and $q=16$ (second
row). Relaxation times are $\protect\tau _{e}=80$ fs, $\protect\tau _{eh}$ =
150 fs. The spot size $\Lambda =1$ $\protect\mu $m, $N_{s}=$10$^{11}$ cm%
$^{-2},$ $U(0)=400$ km/s. }
\end{figure}

This is illustrated in Fig.2, where we show the pattern of the change in the
electron density at two times, for samples with $q=1$ and $q=16$ 
buried deep within bulk Al$_{x}$Ga$_{1-x}$As where Eq.(5) is applicable.
On the time scales of the ballistic
regime, $t\leq\tau _{e},$ in both samples the spot moves as a rigid
distribution with 
\begin{equation}
N_{e}=N_{0}\exp \left[ -\left( x-\left\langle x(t)\right\rangle \right)
^{2}/2\Lambda ^{2}\right] \exp \left[ -y^{2}/2\Lambda ^{2}\right] ,
\end{equation}%
and the corresponding density component develops as $N_{1,0}^{e}(t)=N_{0,0}^{e}(0)\xi/\sqrt{2}$. 
This is clearly seen in the left column of the Fig. 2, where the density
profile is almost the same for systems with weak ($q=1$) and much stronger 
($q=16$) space-charge effects at $t=80$ fs, and is well described by Eq.(8).
At 1.5 ps, the profile for the $q=1$ system buckles only slightly; 
a "rigid spot" picture is applicable. But for $q=16$ the behavior
at these times is much more complicated, with several
$\Psi_{n_{1},n_{2}}$ states contributing, and the profile changing from oval to
bow-shaped. Here the "rigid spot" approximation is
no longer reasonable, and a more general "distorted
puddle" picture emerges.  More insight can be obtained from the current
density $N_{e}(\mathbf{r})u_{x}(\mathbf{r})$ distribution in Fig.3. At $t=80$ fs
the patterns for $q=1$ and $q=16$ look similar. At $t=160$ fs, the
distributions already differ very considerably, and the pattern for $q=16$
is much more nonuniform than that for $q=1$.

\begin{figure}[t]
\includegraphics[width=0.55\columnwidth]{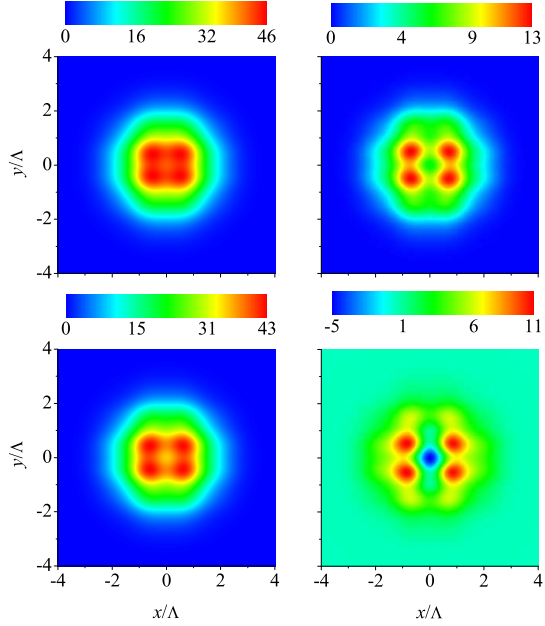}
\caption{(Color online) The electron current density distribution (in arb.
units) at $t=80$ fs (left column) and $t=160$ fs (right column). The MQW
parameters are $q=1$ (upper row) and $q=16$ (second row). Other parameters
are the same as in Fig. 2 }
\end{figure}

It is interesting to see the effect of the real spot shape
on gross features of the density distribution, such as
its first moment (\ref{firstmoment}). We evaluate this from our
calculations as  
\begin{equation}
\left\langle x(t)\right\rangle =\sum_{n_{1}=1}^{n_{%
\lim }}\sum_{n_{2}=0}^{n_{\max }}\frac{N_{n_{1},n_{2}}^{e}(t)}{N_{\mathrm{t}}}\int x\psi
_{n_{1}}\left( x\right) dx\int \psi _{n_{2}}\left( y\right) dy,
\end{equation}%
and present in Fig.4 the results for $q=4$ and $q=16$ samples.  The upper limit 
$n_{\lim }$ here is less than the
number of Hermite-Gaussian states in the basis $n_{\max }$, since the
contributions of the upper states cannot be calculated with a high
precision; this problem is typical for calculations with finite basis. \cite%
{Sherman06a} For this set of parameters we find that $n_{\max }=14$ and $%
n_{\lim }=7$ give converged results. In addition, Fig.4 shows the
displacement in the rigid spot approximation \cite{Sherman06}, where the
dynamics yields $\left\langle x_{\mathrm{rsa}}(t)\right\rangle =
U(0)e^{-\gamma t}\mathrm{sinh}\left({\gamma}_{\rm pl}t\right)/{\gamma}_{\rm pl}$ 
with $2\gamma =1/\tau_{e}+1/2\tau_{eh}$, 
${\gamma}_{\rm pl}^{2}=\gamma^{2}-\Omega_{\mathrm{pl}}^2$.
With the stronger space-charge effects that arise for a larger number of
wells, the deviation from the rigid spot approximation becomes more severe. 
However, both $\left\langle x_{\mathrm{rsa}%
}(t)\right\rangle $ and $\left\langle x(t)\right\rangle $ reach their maxima
at almost the same time 
$t_{0}=2\tau _{e}\left\vert\ln(\Omega _{\mathrm{pl}}\tau _{e})\right\vert$,
and the maximum values are
very close to each other. The
difference begins developing at $t>t_{0}$, when the
buckling effects become more pronounced.  The curves
for small value of $\tau_{eh}=50$ fs show that
a decrease in $\tau_{eh}$ does not qualitatively modify the dynamics
in our range of parameters. However, in a regime with
$\tau_{eh}\ll\tau_{e}$ a decrease in $\tau_{eh}$ can lead to an increase
in the puddle distortion. We turn to the influence 
of  this distortion on experimental results in a separate
publication.  

\begin{figure}[t]
\includegraphics[width=0.30\columnwidth]{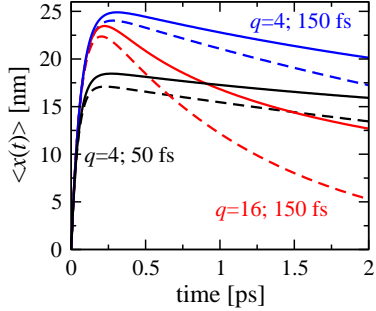}
\caption{(Color online) Mean spot displacement 
$\left\langle x(t)\right\rangle$ (solid lines) and 
$\left\langle x_{\rm rsa}(t)\right\rangle$ (dashed lines). Parameters $q$ and $\tau_{eh}$  
are marked near the lines. Other parameters are the same as in Fig. 2.}
\end{figure}

To conclude, we have shown that within a hydrodynamic model a
Hermite-Gaussian expansion can be used to study the dynamics of currents
injected optically by coherent control in MQW samples. 
As samples with increasing numbers of QWs are considered, we find
the resulting dynamics shows a complicated buckling behavior in the
density distribution that is driven by space-charge effects, with 
the charge density departing more and more from what a rigid
spot approximation would predict. The details of our calculated
results reflect both the simple dynamics (1),(2) we adopt, 
and the assumption of equal excitation of all the quantum wells. But it
is unlikely that a relaxation of these approximations, and the richer
dynamics that would result, would lead to a {\it more rigid} behavior
of the charge puddle. 

For $\tau_{eh}=150$ fs, we find that 
the complex behavior,
associated with a number of comparable Hermite-Gaussian amplitudes,  
emerges at $\Omega_{\rm pl}\tau _{e}\approx 0.1$, well
within the overdamped regime, and is determined by this parameter
rather than by the electron-hole separation. 
How sensitive various experimental probes
will be to this deviation from rigid spot dynamics has yet to be examined;
at least it has significant consequences on the
evolution of the first moment of density on the timescale of a picosecond. 

Our numerical consideration relies on the finite basis analysis. 
In the overdamped regime we have considered here a fast decay of the off-diagonal Coulomb matrix
elements with the distance between the states ensures the applicability of
relatively small basis sets.
As we consider systems with stronger
Coulomb interaction, the basis should be extended.
We can naturally  expect the distortion of the carrier puddles 
to get more pronounced,  and
an interesting problem is whether it can lead to a
chaotic behavior \cite{Pershin07} when many Hermite-Gaussian terms form the density and velocity patterns.

We are grateful to H. van Driel, J. McLeod, A. Najmaie, I. Rumyantsev, and
A.L. Smirl for numerous valuable discussions. This work was supported by the
Natural Sciences and Engineering Research Council of Canada (NSERC).

\end{document}